\documentclass[reprint,amsmath,amssymb,aps]{revtex4-1}
\usepackage{graphicx}
\usepackage{subfigure}
\usepackage{dcolumn}
\usepackage{bm}
\usepackage{amsfonts}
\usepackage{xcolor}
\usepackage{ulem}

\begin{document}
\preprint{APS/123-QED}
\title{Discharge of elongated grains from silo with rotating bottom}

\author{Kiwing To$^1$}
\email{ericto@gate.sinica.edu.tw}
\author{Yi-Kai Mo$^1$}
\author{Tivadar Pong\'o$^{2,3}$}
\author {Tam\'as B\"orzs\"onyi$^2$}

\affiliation{$^1$Institute of Physics, Academia Sinica, Taipei, Taiwan 119 Republic of China}
\affiliation{$^2$Institute for Solid State Physics and Optics, Wigner Research Centre for Physics, P.O. Box 49, H-1525 Budapest, Hungary}
\affiliation{$^3$Departamento de F\'{i}sica y Matem\'{a}tica Aplicada, Facultad de Ciencias, Universidad de Navarra, Pamplona, Spain}

\date{\today}

\begin{abstract}
We study the flow of elongated grains (wooden pegs of length $L$=20 mm with circular cross section of diameter $d_c$=6 and 8 mm) from a silo 
with a rotating bottom and a circular orifice of diameter $D$. 
In the small orifice range ($D/d<5$) clogs are mostly broken by the rotating base, and the flow is intermittent with avalanches and temporary clogs.
Here $d\equiv(\frac{3}{2}d_c^2L)^{1/3}$ is the effective grain diameter.
Unlike for spherical grains, for rods the flow rate $W$ clearly deviates from the power law dependence $W\propto (D-kd)^{2.5}$ at lower orifice sizes in the intermittent regime, where $W$ is measured in between temporary clogs only. Instead, below about $D/d<3$ an exponential dependence $W\propto e^{\kappa D}$ is detected.
Here $k$ and $\kappa$ are constants of order unity. 
Even more importantly, rotating the silo base leads to a strong -- more than 50\% -- decrease of the flow rate, which otherwise does not depend significantly on the value of $\omega$
in the continuous flow regime. In the intermittent regime, $W(\omega)$ appears to follow a non-monotonic trend, although with considerable noise.  
A simple picture, in terms of the switching from funnel flow to mass flow and the alignment of the pegs due to rotation, is proposed to explain the observed difference between spherical and elongated grains.
We also observe shear induced orientational ordering of the pegs at the bottom such that their long axes in average are oriented at a small angle $\langle\theta\rangle \approx 15^\circ$ to the motion of the bottom. 

\end{abstract}

\maketitle

\section{Introduction}
Hoppers and silos are indispensable devices in factories and farms to handle granular materials and grains. Since the orifice of a hopper or a silo is smaller than the main compartment, it becomes a bottleneck when the grains flow out of the appliances.  If the orifice diameter is comparable to the size of the grains, the flow may be clogged permanently by a mechanically stable structure (an arch in two dimensions or a dome in three dimensions) composed of a small number of grains surrounding the orifice. 
Since spherical objects are geometrically simple and conceptually easy to understand, most of the research on silo and hopper flow investigates spherical grains as model systems \cite{To01, To05, Zuriguel03, Corwin08, Hilton10, Zuriguel11, Thomas15, Nicolas18}. 
However, granular materials found in nature and in practical applications (such as tablets and capsules in pharmaceutical industry or agricultural seeds) are often non-spherical. 
With the advances in experimental and computational techniques, there are increasing number of works on the flow dynamics of granular materials composed of non-spherical grains in shear flows \cite{Borzsonyi12, Borzsonyi13, Artoni2019, Campbell2011, Mandal2016, Nagy2017, Trulsson2018, Reddy2009}, hopper flows \cite{Borzsonyi16, Tang16, Ashour17, Szabo18, Vamsi18}, inclined plane flows \cite{Hidalgo2018, Mandal2016, Azema2012} or in rotating drums \cite{Mandal2017}.

For many different kinds of non-spherical particles the relation between the flow rate $W$ and the size (diameter $D$) of the orifice was found to follow Beverloo law: $W\propto \rho \sqrt{g}(D-kd)^{2.5}$ as observed for discharge of spherical grains. Here $\rho$ is the density of the grains, $d$ is a suitably defined effective grain size and $k$ is a number of order unity. One can even compare the effect of geometrical shape to flow rate if suitable parameters for the size and the shape anisotropy of a grain can be defined. For example, the effective grain size $d$ of a rod of length $L$ and circular cross section of diameter $d_c$ can be defined as the diameter of a sphere with the same volume of the grain, i.e., $d=(\frac{3}{2}d_c^2L)^{1/3}$ and the shape-anisotropy can be parameterized by the aspect ratio $L/d_c$. Using these definitions, the Beverloo law for silo flow rate was found to be valid when the aspect ratio is less than six \cite{Ashour17}.

Industrial processes often need to operate with small, but stable flow rates, e.g.~for accurate mixing of different granular components.
With the aim to reach low flow rates and simultaneously prevent/avoid clogging during the operation of the silo, there has been a lot of research effort to understand the physics of the flow and clog phenomena during the discharge of granular materials from silos and hoppers. In practice one can use external means (such as vibration, air jets, ... etc) to release the clog when it occurs. Recently \cite{To19, To17}, it was found that small motion of the orifice or rotation of the silo base could be an effective way to prevent clogging during discharge of mono-disperse spherical beads from a two-dimensional (2D) or three-dimensional (3D) silo. 

Surprisingly, while the motion of the orifice enhances the flow rate when the orifice size is small, it reduces the flow rate at large orifice size in the 2D silo \cite{To17}. Similarly, rotation of the silo base enhances the flow rate at small orifice sizes but leads to a complex non-monotonic behavior of the flow rate at large orifice sizes in 3D silos \cite{To19}.
This interesting phenomenon has recently been investigated numerically by Hern\'andez-Delfin et.~al.~\cite{Delfin20} who have confirmed the emergence of a horizontal current at the bottom due to the rotation of the bottom of the silo with respect to the wall. 
Increasing the rotation rate $\omega$ induces a complex change in the velocity field and packing fraction of the material, leading to a non-monotonic (first decreasing then increasing) flow rate with $\omega$.

It is not only practically useful to test whether the effects of bottom rotation to the flow rate for spherical grains are also observed for non-spherical grains, but also interesting to understand how the shape anisotropy of the grains affect the static and dynamic properties of the grains near a bottle neck.
For example, using x-ray tomography, B\"orzs\"onyi et.~al.~\cite{Borzsonyi16} showed clearly the ordering and alignment of the elongated grains in the static packing near the orifice of a clogged silo before and after an avalanche.
This result suggests that the rotational degree of freedom of elongated grains is coupled to the flow field associated with the discharge process. 
When the flow field is affected by external means, the behavior of the elongated grains will be different from that of the spherical grains. 

In this paper we report our experimental results of the discharge of elongated grains from a silo with a rotating bottom. 
We find the same feature that permanent clogging in silos of small orifice size can be prevented by rotating the bottom for elongated grains as for the spherical grains. 
However, for larger orifice sizes the flow rate for elongated grains is much more affected by the rotating bottom than for spherical particles. Namely, for elongated grains we observe a reduction of the flow rate by  more than 50\%.
This surprising result may be explained by the motion of the pegs at the bottom towards the orifice and the orientation ordering induced by the rotation of the bottom. 

In the following section, we describe our experimental setup and explain the procedures for measuring the flow rate in different flow regimes. Then our findings and our understanding of the variations of the flow rate with orifice size, rotation speed and position of orifice as well as the alignment of pegs at the bottom are presented in section III. A summary containing the main findings is given at the end.

\section{Setup and Procedures}
\begin{figure}[t]  \centering \includegraphics[width=8cm]{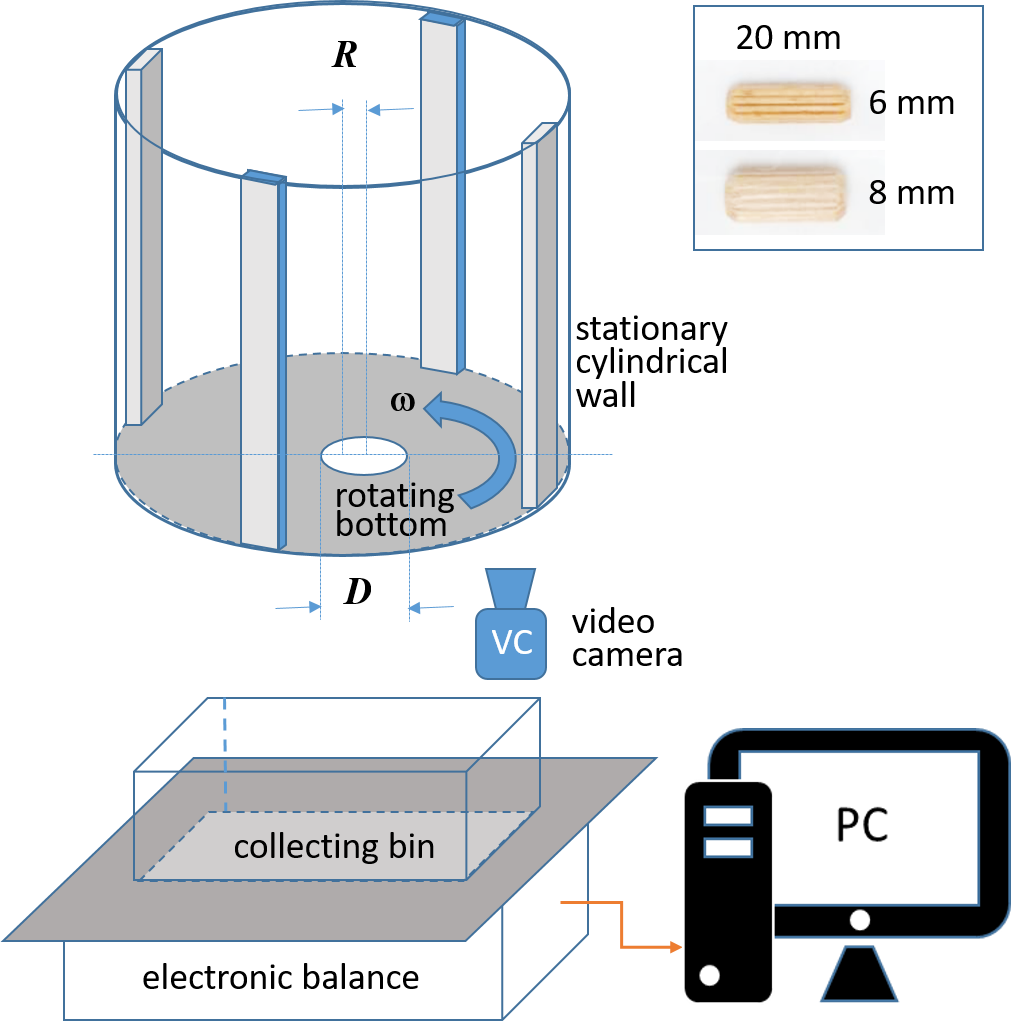}\caption{Schematic diagram of the experimental setup. In the upper right corner photos of the thin and thick pegs are shown.}  \label{fig:setup}\end{figure}

The experimental setup used in our studies is similar to that in Ref.~\cite{To19}. 
The schematic diagram is shown in FIG. \ref{fig:setup}.
It consists of a 19 cm inner diameter cylindrical acrylic silo with a rotatable bottom in which a circular orifice of diameter $D$ is cut at a distance $R$ from the axis of the silo. 
Four vertical plastic bars of 3 mm thickness and 18 mm width are securely glued on the cylindrical wall of the silo to prevent the granular packing inside the silo from performing solid rotation when driven by the rotating bottom of the silo.  
An electronic balance (Sartorius MSE36201S, 10$^{-4}$ kg resolution) with a plastic bin is placed below the bottom of the silo to collect the material discharged from the silo.
The bottom of the silo is driven by a DC motor (Oriental Motor BLFD30A2) via a belt and its rotation speed $\omega$ can be controlled from 0.01 revolutions per second (rps) up to 1.50 rps.
A video camera (VC) is mounted below the silo for taking pictures of the grains above the transparent bottom plate.

Two types of elongated grains of aspect ratios 2.5 and 3.3 are used in our studies. 
They are 20 mm long wooden pegs with circular cross section and grooves on their surface.   
The diameter of the cross section for the thin (and thick) pegs is $d_c$=6 mm (and $d_c$=8 mm) and hence, its effective grain size is $d$=10.26 mm (and 12.43 mm). 
The masses of a thin and a thick peg are respectively 0.35 and 0.52 g. 
Thus, the shape of both types of elongated grains strongly deviates from a sphere, and as we will see, the main findings of our experiments are the same for both of them.
The following is a description of the procedures to measure the flow rate of the pegs from the silo.

At the beginning of an experiment pegs are loaded into the silo with the orifice blocked. 
For all experiments the same filling procedure was applied, resulting in an initial packing fraction with minor ($<2\%$) variation. In principle, a strong difference in the initial orientation of the grains could lead to different discharge scenarios, which will be the subject of future research.
After filling, we set the bottom of the silo to rotate, reset the electronic balance to zero and remove the block from the orifice. 
The pegs from the silo fall out of the orifice into the collecting bin on the electronic balance which sends the measured mass $m$ to a personal computer (PC: Asus Eee BOX B202) at a data rate of 10 readings per second. 
In our experiments, data in the range 0.01 kg $< m <$ 1.5 kg are used for discharge flow rate measurement to avoid the initial transient due to unblocking the orifice, as well as the non-linear regime at the end of the discharge process.
For each combination of the control parameters $(D,\omega)$, the mean discharge rate is calculated from at least five independent measurements.

\section{results and discussion}
\subsection{Flow behavior}

\begin{figure}[t]  \includegraphics[width=\linewidth]{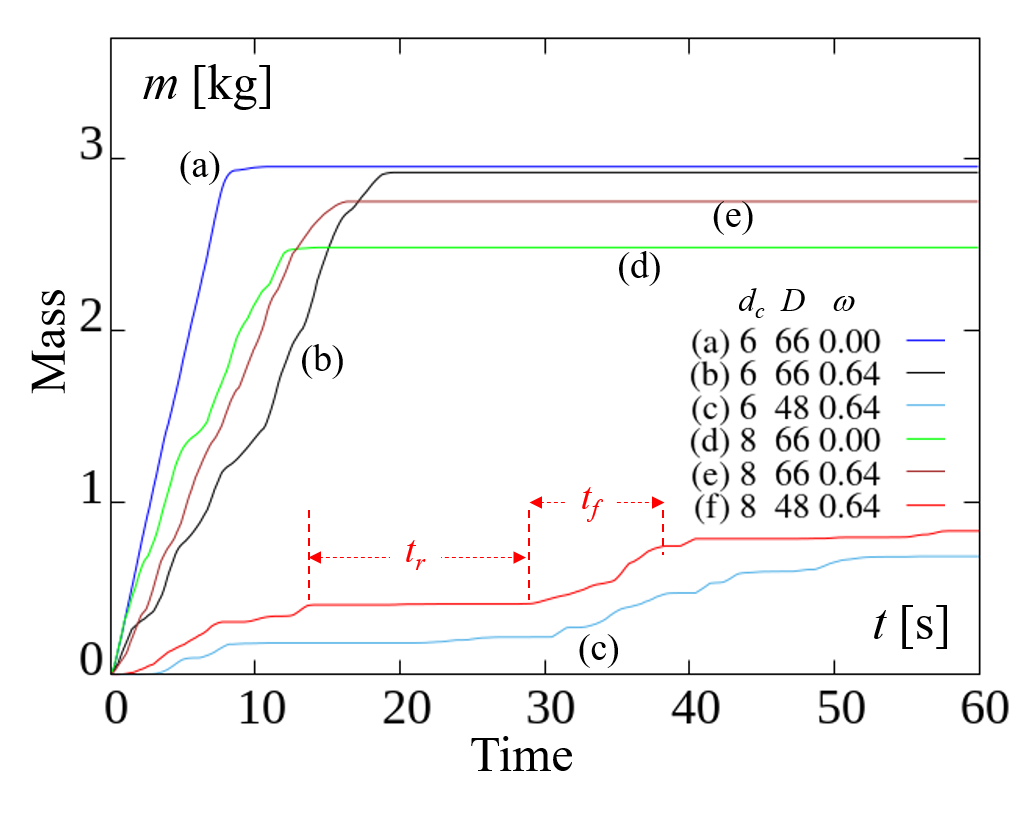} \caption{Time evolution of the reading $m$ from the electronic balance for silo with two different orifice diameters ($D$ in mm) and rotation speeds ($\omega$ in rps) for the thin ($d_c$=6 mm) and thick ($d_c$=8 mm) pegs.}   \label{fig:rawdata}\end{figure}

Figure \ref{fig:rawdata} shows six data sets of the temporal evolution of the mass $m$ collected at different combinations of peg type $d_c$, orifice size $D$ and rotation speed $\omega$ with the orifice positioned at the center of the silo (i.e. $R$ = 0). 
For the non-rotating ($\omega$ = 0) silo with a large orifice ($D$ = 66 mm), it takes less than 10 s to discharge 3 kg of the thin pegs ($d_c$ = 6 mm) as shown by the data set (a) in the figure. If the bottom of the silo rotates at 0.64 rps, it takes almost 20 s to empty the silo (see data set (b)). 
For a smaller orifice ($D$ = 48 mm) rotating at the same speed (data set (c)), temporary clogs (identified as time intervals of 1 s or longer in which $m$ remains unchanged) are observed during the discharge process. If the bottom does not rotate, persistent clog, defined as no change of $m$ within 5 minutes, occurs during discharge of thin pegs for orifice of diameter less than 54 mm. 

These three kinds of flow behavior: continuous flow, intermittent flow and persistent clog are also observed in the discharge of thick pegs ($d_c$ = 8 mm). For non-rotating silo base, persistent clog occurs for the thick pegs when $D <$ 60 mm.
At the largest orifice diameter ($D$ = 66 mm) in our experiments, the flow is continuous as illustrated by data set (d) in FIG. \ref{fig:rawdata}.
When $D$ is reduced to 48 mm, the flow becomes intermittent (see data set (f) in FIG. \ref{fig:rawdata}).

During a temporary clog, for example, beginning from $t$ equals 14.0 s in data set (f) of FIG. \ref{fig:rawdata}, the configuration of the packing above the orifice was slowly changing due to the motion of the bottom plate. After $t_r$=15.1 s, the packing reached an unstable configuration and the clog was released at time equaling 29.1 s. Pegs were then discharged from the silo for $t_f$=9.4 s. Then at 38.5 s the flow was clogged again. 
One can define the flow time $t_f$ as the time interval between two consecutive clogging events and the recovery time $t_r$ as the time interval between two consecutive flowing events as shown in the figure.
The mean flow time and the mean recovery time are found to increase and decrease, respectively, with the rotation speed.
Due to the finite response time of the electronic balance which has a settling time of 1 s, an event such that $m$ does not change in a time interval shorter than 1 s is not considered as an intermittent clog.

\begin{figure}[t]  \includegraphics[width=\linewidth]{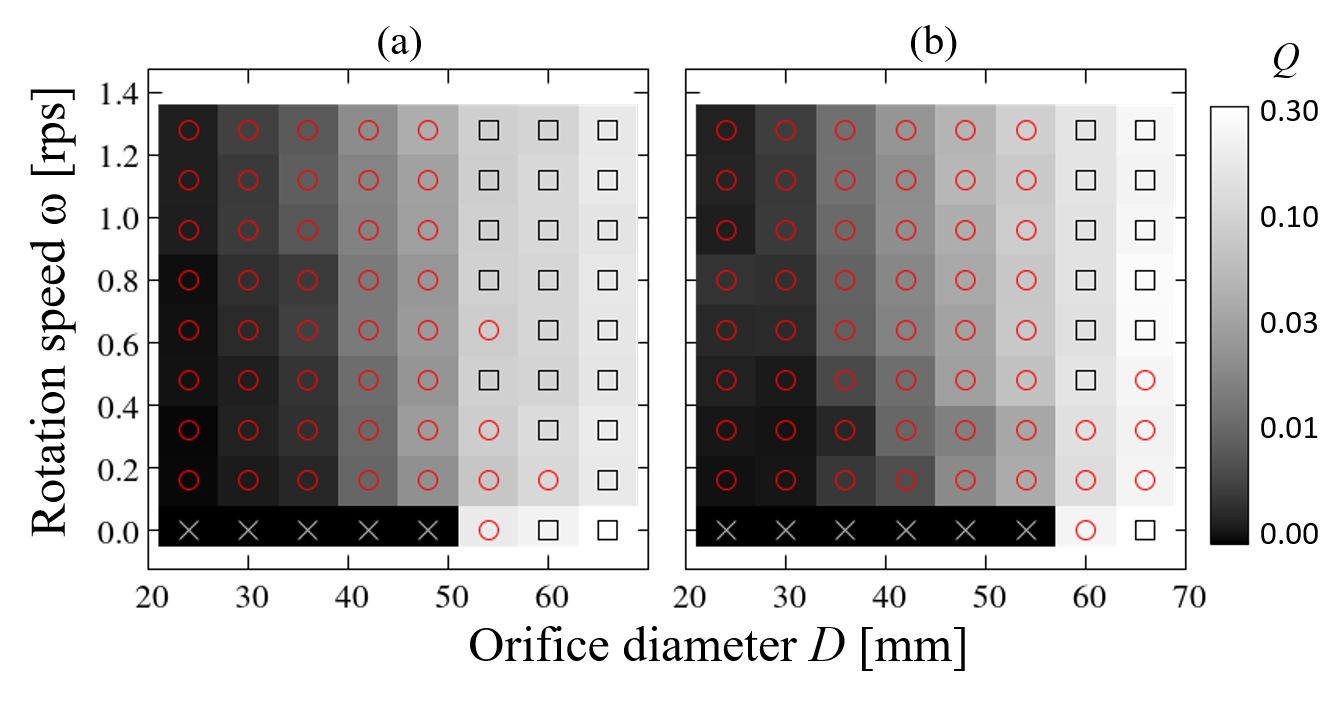} \caption{Discharge flow regimes for thin (a) and thick (b) pegs at various values of the orifice diameter $D$ and rotation speed $\omega$. Continuous flow: $\square$, intermittent flow: \textcolor{red}{$\bigcirc$}, persistent clog: $\times$. The value of the average flow rate $Q$ in kg/s is indicated by the grey scale background behind the symbols. }  \label{fig:flowtype}\end{figure}

The conditions, in terms of orifice diameter and rotation speed, for persistent clog, intermittent flow and continuous flow are summarized in FIG.~\ref{fig:flowtype}.
In all of our experiments, rotating the base serves to eliminate persistent clogs.
When discharging thin pegs, no continuous flow is observed for $D<$ 54 mm. 
For the case of $D$ = 54 mm continuous flow can be maintained only with fast rotation of the silo base.
For larger orifice diameter $D\geq$ 60 mm, the flow is continuous except for the slowest rotation speed $\omega$ = 0.16 rps at $D=$ 60 mm.
For the thick pegs, which have a larger effective diameter than that of the thin pegs, the transition from intermittent flow to continuous flow moves to larger orifice diameter ($D\gtrsim$ 60 mm) and faster rotation speed ($\omega\gtrsim$ 0.48 rps). 
While the grey scale background behind the symbols in FIG. 3 indicates the average flow rate which vanishes in the persistent clog regime, there is no sharp change in the average flow rate from intermittent to continuous flow regimes.

\subsection{Average and avalanche flow rates}
Intermittent flow behavior has been observed in tilted silo \cite{Thomas13}, silo under vertical vibration \cite{Janda08,Mankoc07} and two-dimensional silo with an oscillating exit \cite{To17}. 
On the one hand, one can measure the average flow rate $Q=\Delta m/\Delta t$ by taking the ratio between the mass discharged $\Delta m$ and the time interval $\Delta t$ of measurement which includes all flow time and recovery time, i.e. $\Delta t=\sum{(t_f+t_r)}$.
Thus, $Q$ also involves the temporary clogs, and is important from a practical point of view.
According to ref.~\cite{To17}, $t_r$ and $t_f$ follow different statistics. 
Furthermore, the average flow rate does not obey Beverloo law in the intermittent flow regime. 

On the other hand, it is of fundamental interest to define an avalanche (or active) flow rate $W=\Delta m/\sum{t_f}$ as the mass of the discharged material $\Delta m$ divided by the total flow time only (i.e. excluding temporary clogs). This quantity is relevant in analyzing how the flow dynamics is changing across the transition from continuous flow to the clogging regime  with decreasing orifice size. Interestingly, several investigations indicate that there is no sharp transition in avalanche flow rate, and Beverloo law is still valid in the intermittent and clogging regime \cite{Mankoc07,Janda08,Thomas13,To17}.

The avalanche flow rate equals the average flow rate in the continuous flow regime because in this case the recovery time vanishes. 
To measure the avalanche flow rate in the presence of persistent clog, we wait for a time interval $t_w > 3$ s to make sure that the clog is indeed persistent and then we poke a stick through the orifice to release the blockage. Since a finite amount of pegs falls out of the silo in finite time interval, an avalanche flow rate similar to that of intermittent flow can be measured with $t_w$ playing the role of $t_r$.

\begin{figure}[t] \includegraphics[width=\linewidth]{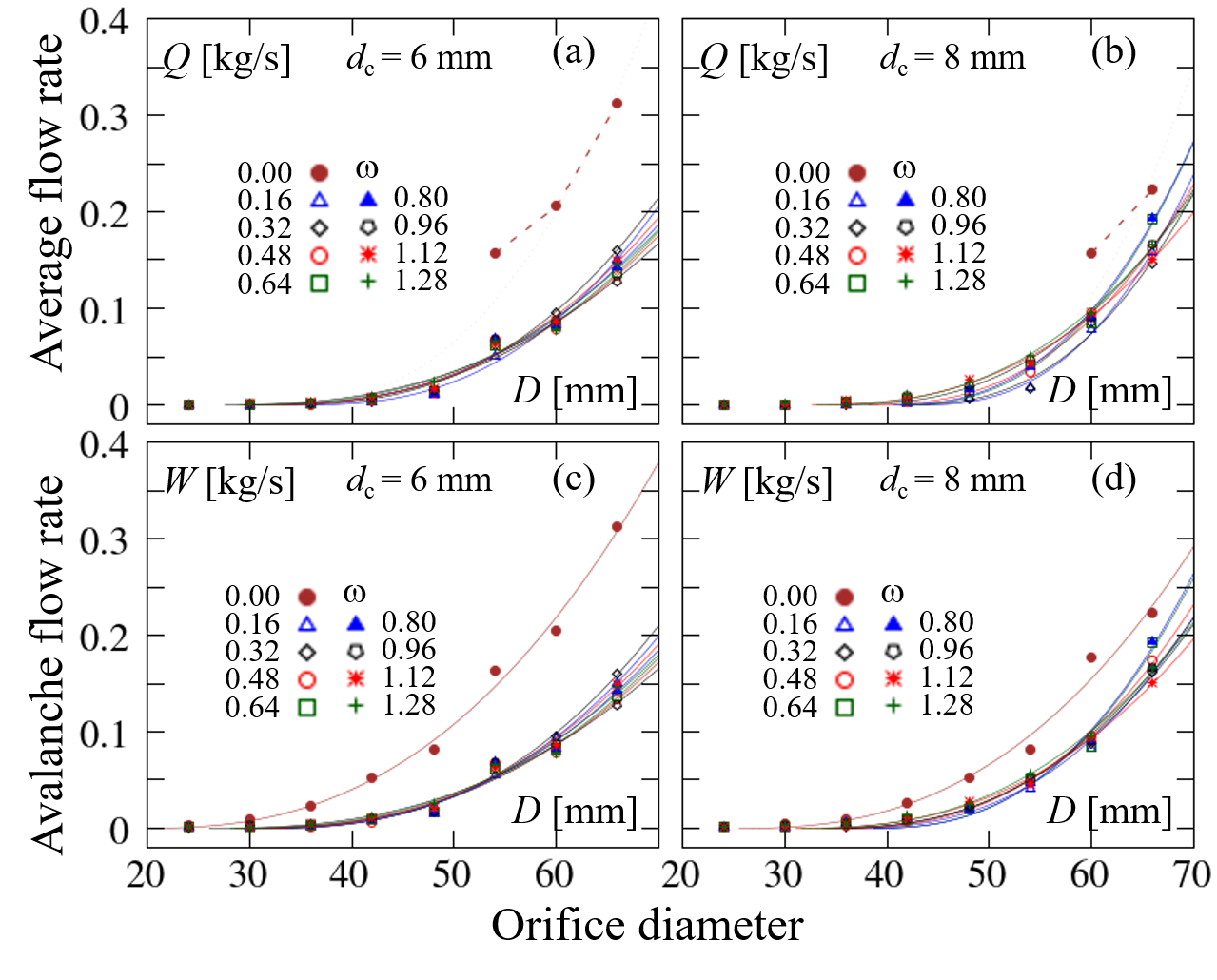}
\caption{Average flow rate $Q$ vs orifice diameter $D$ for (a) thin pegs with $d_c=$ 6 mm and (b) thick pegs with $d_c=$ 8 mm. Avalanche flow rate $W$ for (c) thin pegs and (d) thick pegs vs orifice diameter. Lines in the plots are attempted fitting curves to Beverloo law. The dashed lines in (a) and (b) are guide to the eye only. } \label{fig:BeverlooLinear}\end{figure}

\begin{figure}[t]\includegraphics[width=\linewidth]{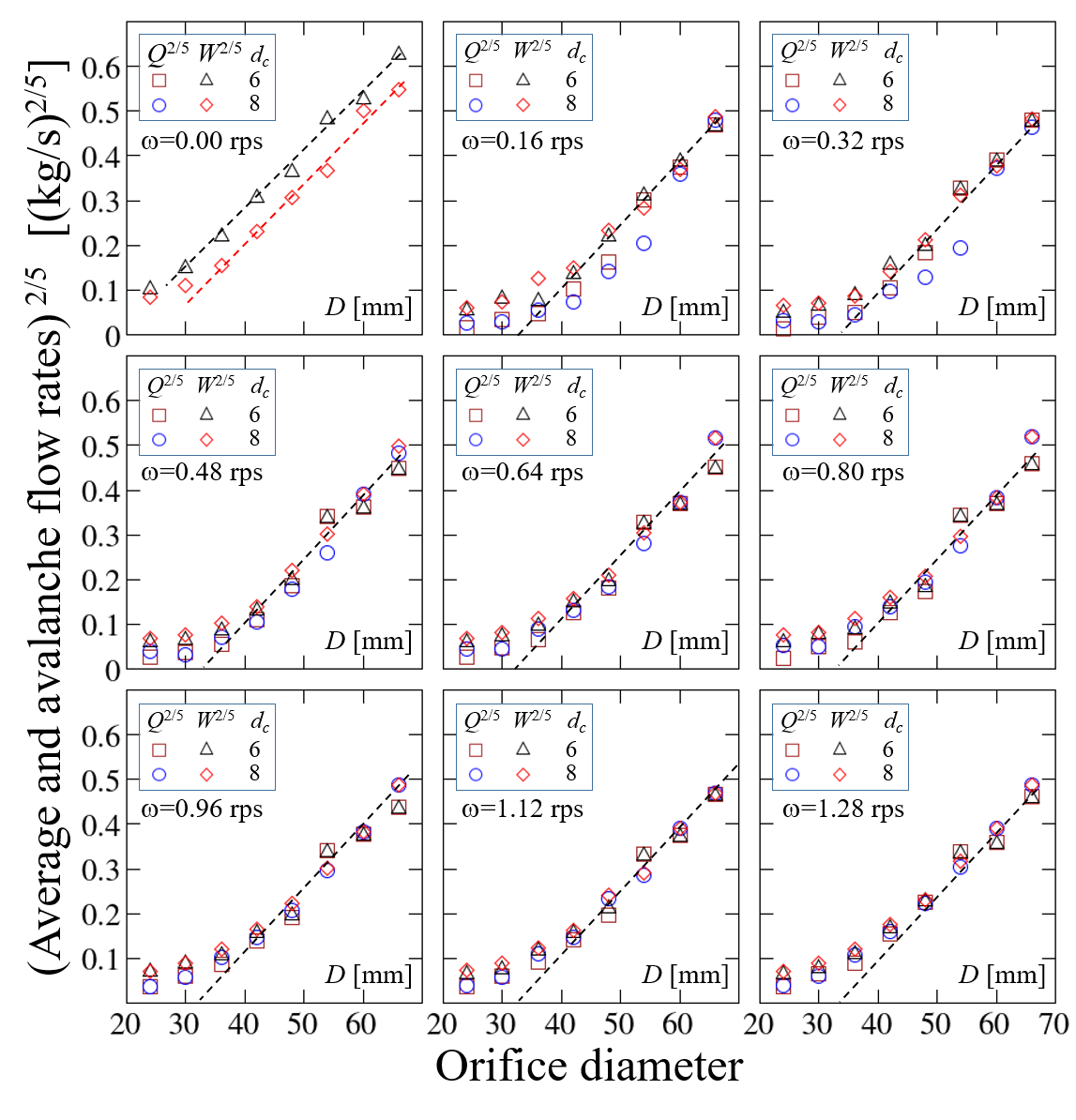}\caption{ Average flow rate $Q$ and avalanche flow rate $W$ vs orifice diameter $D$ plotted in two-fifth power for different rotation speeds $\omega$ = 0.00, 0.16, ..., 1.12 and 1.28 rps. Dashed lines in the plots are guides to the eyes for comparison to Beverloo law. } \label{fig:BeverlooPower} \end{figure}

\begin{figure}[t]\includegraphics[width=\linewidth]{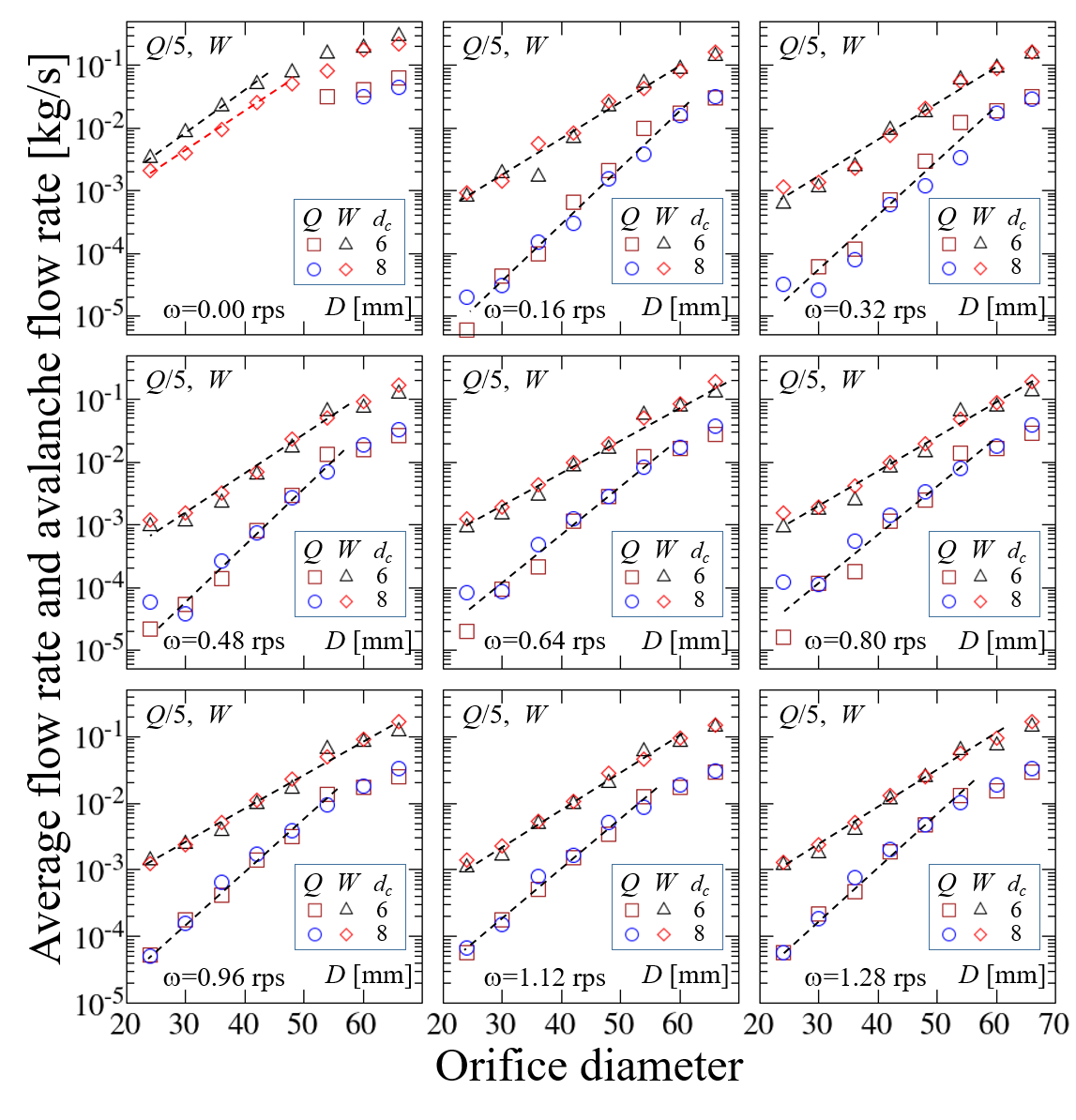}\caption{Average flow rate $Q$ and avalanche flow rate $W$ vs orifice diameter $D$ for different rotation speeds $\omega = 0.00, 0.16, \cdots, 1.12$ and 1.28 rps. For better viewing of the data, the values of $Q$ in the plots are divided by 5. Dashed lines in the plots are guides to the eyes for comparison to exponential law.} \label{fig:Semi-LogW} \end{figure}

FIG. \ref{fig:BeverlooLinear}(a) and (b) show, respectively, the average flow rate $Q$ for thin ($d_c=$6 mm) and thick ($d_c=$ 8 mm) pegs versus the orifice size $D$ for different rotation speeds $\omega = 0, 0.16, \cdots, 1.12$ and 1.28 rps. 
The avalanche flow rate $W$ for the same rotation speeds is plotted in FIG. \ref{fig:BeverlooLinear}(c) and (d).
The lines in these plots are attempted fitting curves to Beverloo law:  $Q$ or $W = C\rho\sqrt{g}(D-kd)^{2.5}$.
When the bottom of the silo is not rotating and $D$ is small, $Q$ vanishes due to permanent clogging and hence, we only have three data points for the thin pegs and two data points for the thick pegs. On the other hand, $W$ at $\omega$ = 0 appears to follow Beverloo law down to the smallest orifice diameter. 
At finite rotation rate, both $Q$ and $W$ seem to fit Beverloo law reasonably well.
However, Beverloo law is valid only for large orifice diameters.
This is better illustrated in FIG. \ref{fig:BeverlooPower} in which $Q^{2/5}$ and $W^{2/5}$, respectively, are plotted against $D$.
If a data set obeys Beverloo law, it should fall on a straight line in the plot.

One can see from the top left plot in FIG. \ref{fig:BeverlooPower} that $W$ can be fitted to Beverloo law when $\omega=0$. Deviation from Beverloo law happens when $D\lesssim 30$ mm and $D\lesssim$ 36 mm for the thin and thick pegs, respectively. 
These numbers set the lower bound of the orifice diameter to $D/d\approx 2.9$ times the diameter of the effective grain size $d$ for the validity of Beverloo law.   
In silos with rotating bottom, both $Q$ and $W$ fit Beverloo law reasonably well only when $D\gtrsim$ 40 mm.
 
While the validity of Beverloo law for spherical grains in a silo with rotating bottom \cite{To19} as well as for elongated grains in a silo with stationary orifice \cite{Ashour17} have been verified, the results presented here suggest that the rotation of the orifice does not affect the validity of Beverloo law for elongated grains even when crossing the boundary between continuous flow and intermittent flow.

Deviations of the avalanche flow rate from Beverloo law have also been reported for spherical grains in the small orifice (clogging) regime \cite{Mankoc07}. In that study, an exponential correction term was introduced for the small $D$ range and the physical origin of the correction was shown to be related to the spatial self-similar nature of the density and velocity profiles close to the orifice \cite{Janda12}.
Interestingly, when we plot the flow rates of the thin and thick pegs versus orifice diameter in semi-log graphs, the exponential relation between the flow rates and orifice diameter is revealed for silo with rotating bottom in the intermittent regime down to the smallest orifice as shown in FIG. \ref{fig:Semi-LogW}.
In addition, at fast rotation speeds ($\omega >$ 0.9 rps) the avalanche flow rates $W$ for the thin and the thick pegs are the same within experimental uncertainty and they can be fitted to: $W=W_o  e^{\kappa D}$ with $W_o=5.0\times10^{-5}$ kg/s and $\kappa=0.13$ mm$^{-1}$. 
The average flow rates $Q$ have similar behavior and they can be fitted to: $Q=Q_o  e^{\kappa D}$ with $Q_o=4.5\times10^{-6}$ kg/s and $\kappa=0.18$ mm$^{-1}$.
More effort is needed to understand the physical origin of such dependence.

\subsection{Effects of rotation on flow rate}

The failure of Beverloo law in the intermittent flow regime for elongated grains due to the rotation of the bottom suggests that the concept of free fall arch may not be applicable in this situation.
In the free fall arch model, the current through the orifice is calculated by the vertical motion of the grains while the horizontal flux along the bottom is ignored.
Then it is mainly those grains above the orifice that contribute to the discharge current.
This assumption may be justified in funnel flow due to the presence of a stagnant zone at the bottom of the silo.
If there is significant contribution to the discharge from the horizontal current along the bottom, the free fall arch argument will not be applicable and Beverloo law may be violated.
This can happen in our experiments, even though our grains are non-spherical, if the bottom rotation is fast enough to fluidize the grains close to the bottom and switch the dynamics from funnel flow to mass flow. 

\begin{figure}[t] \includegraphics[width=\linewidth]{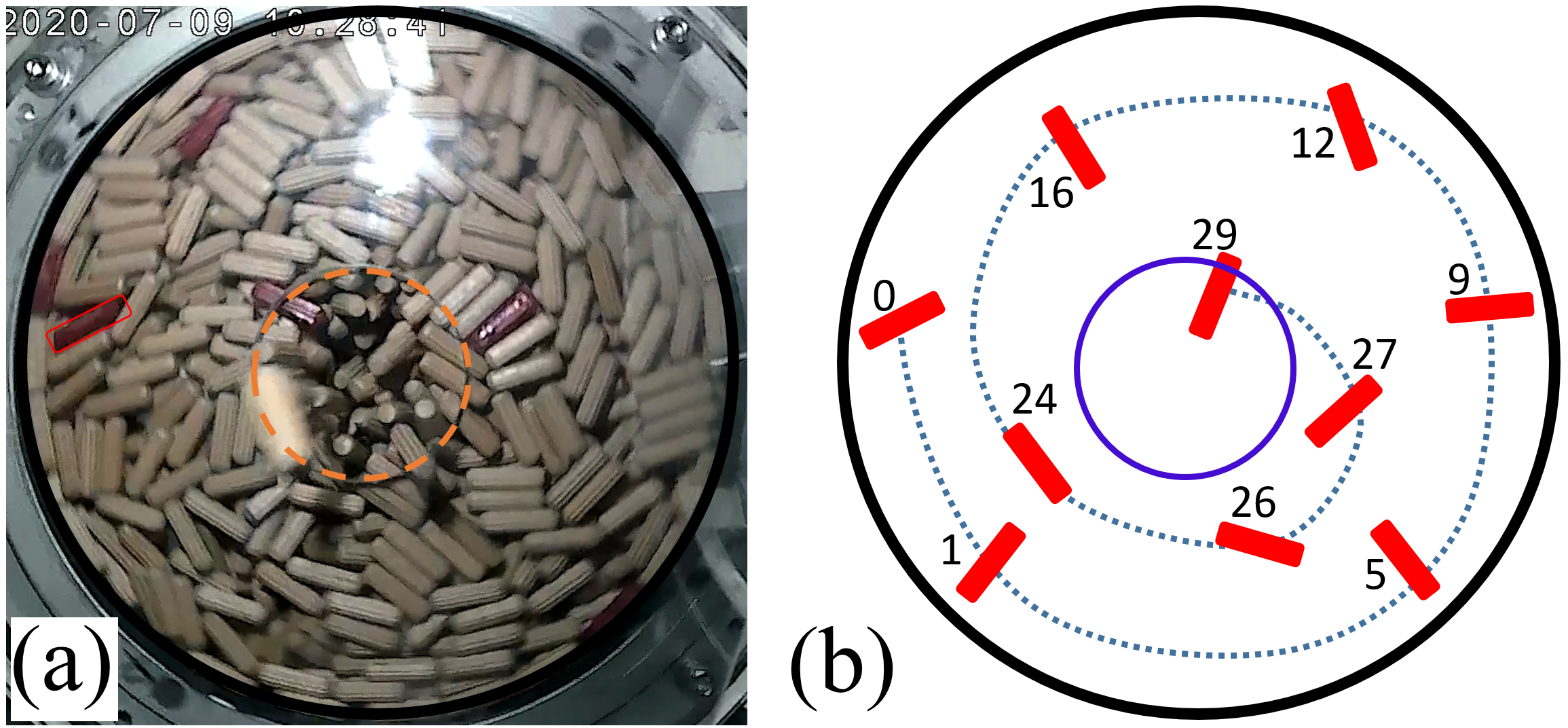} \caption{(a) Photograph of the silo taken from below. The orifice with diameter $D$ = 60 mm is indicated with orange dashed line, the rotation rate was $\omega=0.48$ rps. The motion of particles in the bottom layer is tracked by red tracer particles.  The trajectory of a tracer peg (outlined by a red rectangle) is shown in (b) with red bars indicating the positions and orientations of the peg at later times. The numbers next to the bars indicate the time, in second, relative to that of (a).} \label{fig:PegTraj} \end{figure}

Therefore, we watch the motion of the pegs through the cylindrical wall under rotation and we find that some pegs move downwards to the bottom, then move inwards to the center and finally fall out through the orifice. On the contrary, pegs at the bottom remain stationary until most of the pegs are discharged. 
FIG. \ref{fig:PegTraj}(a) shows a picture from a video taken by the camera below the silo with $D$ = 60 mm at $\omega$ = 0.48 rps. In that particular experiment, a small number of pegs were painted red for better observation of their orientation on their way to the orifice. The pegs next to the cylindrical wall are mostly aligned with their long axes parallel to the wall, i.e. perpendicular to the radial direction. Nevertheless, a peg changes its orientation when it moves towards the orifice as shown in FIG. \ref{fig:PegTraj}(b). The alignment of the pegs at the bottom is discussed in Sec.~\ref{sec:alignment}.

\begin{figure}[t]  \raggedleft  \includegraphics[width=\linewidth]{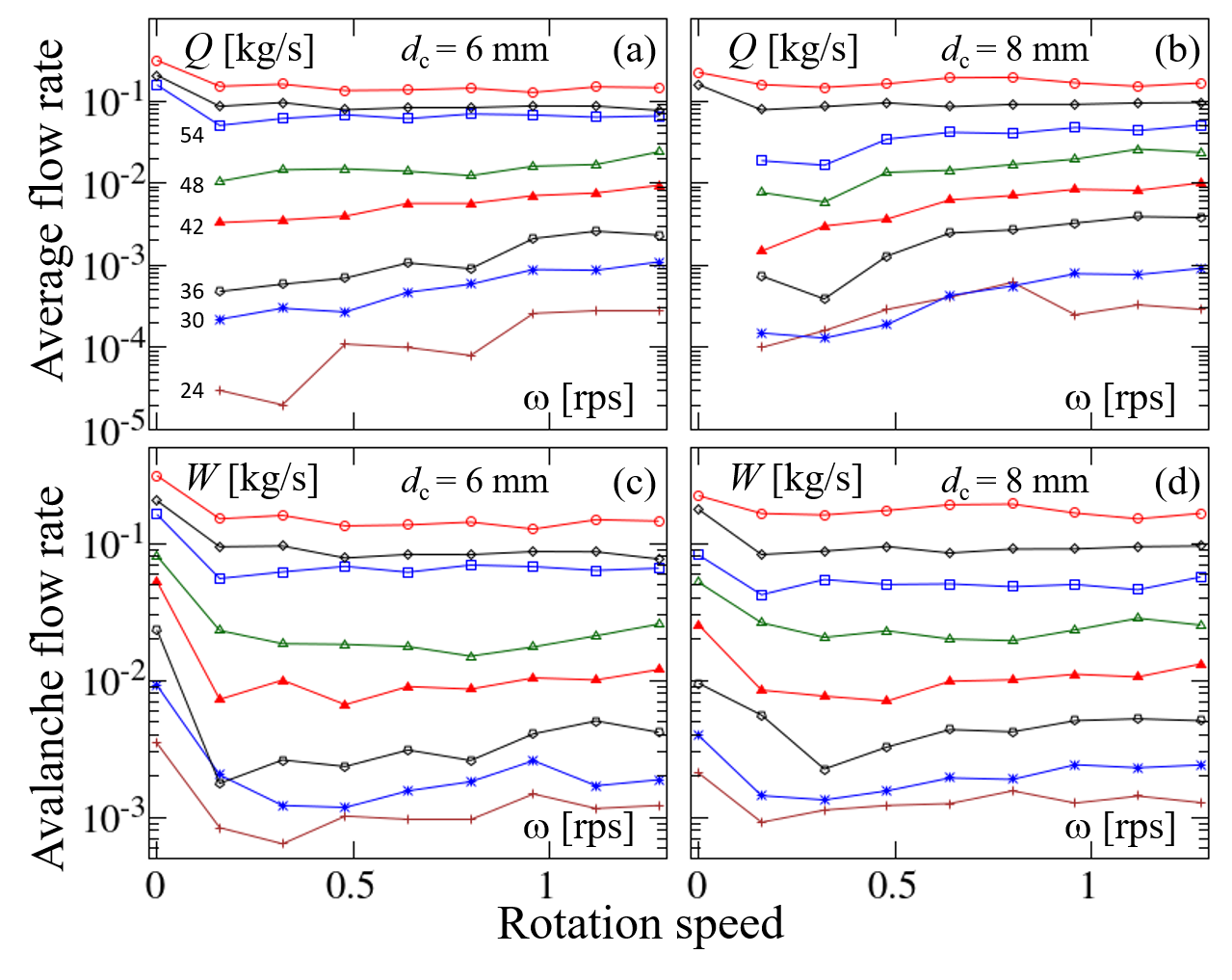}\caption{Average flow rate $Q$ and avalanche flow rate $W$ vs rotation speed $\omega$ for orifice diameter $D$ = 24, 30, $\cdots$, 60 (black diamond) and 66 (red circle) mm of the thin and thick pegs. The numbers in (a) close to the data sets are the values of $D$ in mm and these values also apply to (b), (c) and (d). }  \label{fig:p68fQW-all}\end{figure}

When we plot the average and avalanche flow rate against the rotation speed $\omega$ (see FIG. \ref{fig:p68fQW-all}), we find that the average flow rate $Q$ of the silos with rotating bottom are always smaller than those of the silos with stationary bottom in the continuous flow regimes.
Increasing the rotation rate leads to an exponential increase in $Q$ in the intermittent flow regimes. The growth rate of this exponential increase decreases with orifice diameter. At large orifice sizes (approximately $D>50$ mm) the actual value of the rotation rate has less effect on the flow rate. Similar behavior of the dependence on rotation speed is observed for avalanche flow rate $W$.
Nevertheless, the drop in $W$ when $\omega$ changes from 0 to 0.16 rps is larger than that in $Q$. 
Furthermore, enhancement of $W$ due to increasing $\omega$ is less obvious than that of $Q$.

It is worth comparing the discharge of spherical grains in similar situations, in which the flow rate was either increasing (in the intermittent flow regime at small orifice sizes) or non-monotonic (in the continuous flow regime at larger orifice sizes) with increasing rotation rate \cite{To19}. Furthermore, in the continuous flow regime the relative variation of the first decreasing and then increasing trend in flow rate was less than 10\%. In our experiments with elongated grains a strongly decreasing trend is observed, the flow rate decreases to less than 50\% at high rotation rate compared to its value without rotation. To explain this remarkably strong difference in the dependence of the flow rate on the rotation of the bottom for spherical and elongated grains, we need to examine the effects of rotation on the granular packing in the silo.

Rotation of the bottom relative to the cylindrical wall can be considered as perturbations to the contact forces among the grains inside the silo. 
When the grains are flowing out through the orifice, the fluctuating contact forces among the grains make them more difficult to form a mechanically stable structure (a dome) at the orifice to block the flow.
Even when the orifice is clogged, these perturbations can destabilize the dome and release the clog. 
These are the two main reasons for the flow rate enhancement by rotation for spherical grains \cite{To17,To19}. These arguments are also valid for elongated grains and they can explain our results in the intermittent flow regime. 

In a continuous flow regime stable arches and domes are no longer present. 
Nevertheless, the motion of the bottom plate may switch the discharge process from funnel flow to mass flow as mentioned before. This indeed happens in our experiments. From video images of the pegs at the bottom of the silo, we find that these pegs are stagnant and pegs in the silo are discharged in funnel flow if the bottom does not rotate. 
Finite rotation of the bottom plate turns the discharge process to mass flow because pegs at bottom are observed to move to the orifice, as shown in FIG. \ref{fig:PegTraj}(b), and contribute to the total discharge current. 
These pegs, unlike those in the downward current falling directly through the orifice, have no downward velocity. They become obstacles to the current flowing along the axis of the silo.
This is the scenario confirmed in previous studies on silo discharge of spherical grains \cite{To19,Delfin20}. 
Here the axes of the pegs at the bottom are mostly horizontal while those pegs in the central current are aligned with their axes vertical. Hence, the pegs from the bottom at the orifice have a larger cross section in the vertical direction than those from above the orifice and the reduction of the discharge flow rate of pegs is much larger than that of spherical grains. 

\subsection{Pegs alignment}\label{sec:alignment}
The axes of the pegs at the bottom plate is aligned horizontally due to the boundary effect. It is interesting to see if these pegs are further aligned by the motion of the bottom plate. 
Rotation of the bottom plate imposes shear stress to the grain packing \cite{Corwin08,Hilton10}.
While shearing on a spherical grain has no effect on its orientation, an elongated grain will tend to align its long axis nearly parallel to the flow (i.e. nearly perpendicular to the velocity gradient) \cite{Borzsonyi12,Borzsonyi13}. Without rotating the bottom plate, silo flow already involves shear, which leads to gradual ordering of the elongated grains as they sink in the silo. When they approach the exit, the average alignment of their long axes points nearly towards the orifice \cite{Borzsonyi16}. 
In our experiments in which the bottom plate rotates, the long axis of the pegs at the bottom is expected to be aligned approximately perpendicular to the radial direction. 

\begin{figure}[t]  \includegraphics[width=\linewidth]{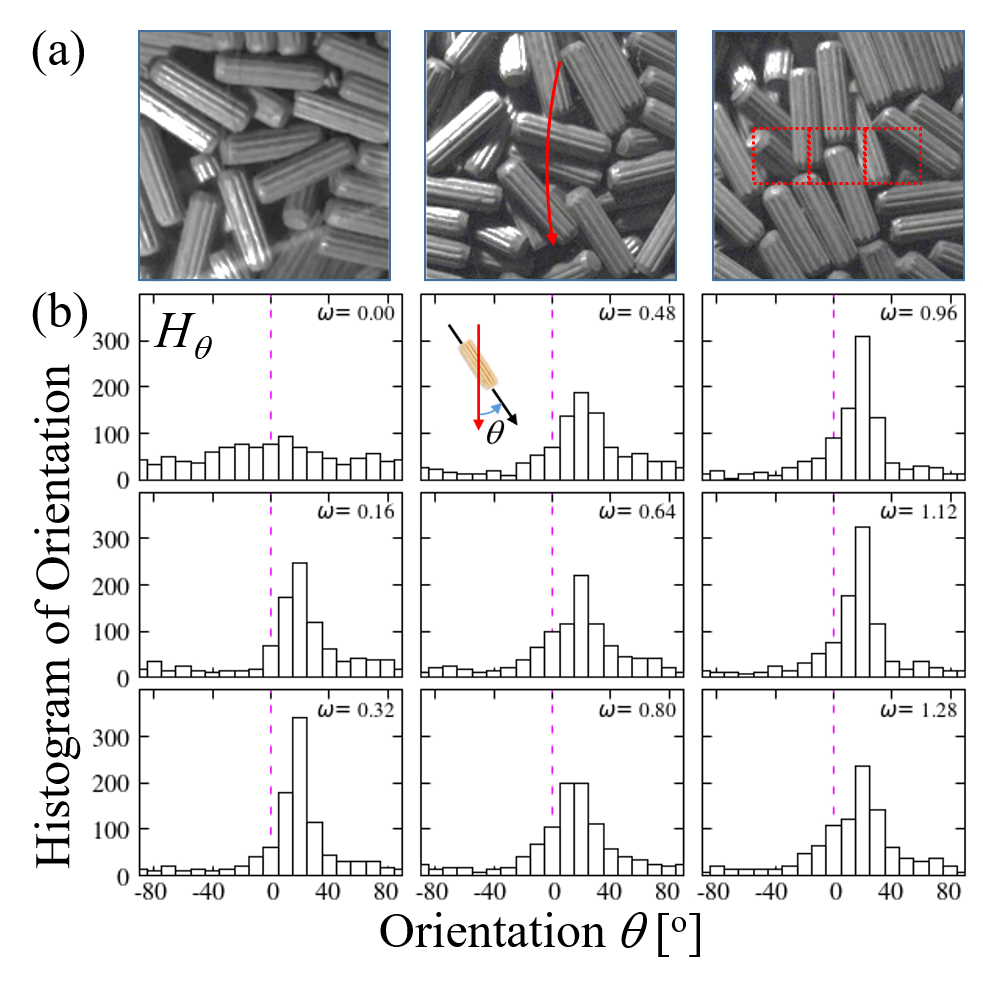}  \caption{(a) Images of the thin pegs taken by the camera at the bottom of the silo of orifice diameter $D=54$ mm at rotation speed (from left to right) $\omega$ = 0.00 rps, 0.48 rps and 0.96 rps. The red arrow in the middle image shows the direction of the motion of the bottom and the dotted squares in the right image are the regions where the orientation angle is sampled. (b) Histograms $H_\theta$ of the orientation angle $\theta$ enclosed by the long axis of the peg and the direction of motion of the bottom. The value of $\omega$ in rps is indicated in the histograms. The schematic diagram in the histogram of $\omega=0.48$ rps defines the orientation angle $\theta$ which is the angle between the long axis of a peg (black arrow) with respect to the motion of the bottom (red arrow).} \label{fig:HistThetaAll}\end{figure} 

\begin{figure}[t] \includegraphics[width=\linewidth]{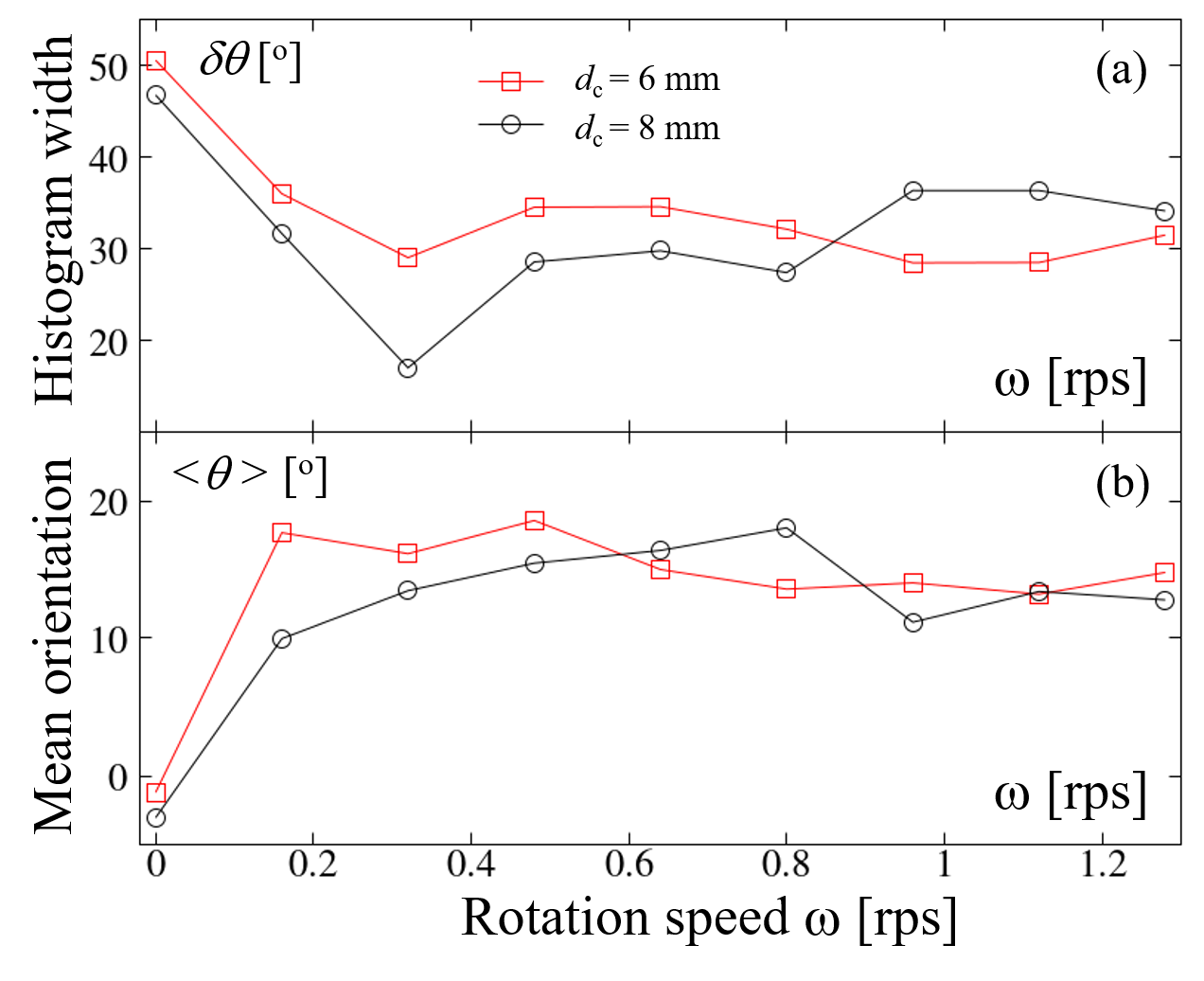} \caption{(a) The width of the orientation angle distribution $\delta\theta$ and (b) the mean orientation angle $\langle\theta\rangle$ vs rotation speed $\omega$ for the thin and thick pegs. } \label{fig:MeanTheta} \end{figure}

To check if the above speculation is true, we examine the images taken from the camera at the bottom of the silo.  FIG. \ref{fig:HistThetaAll}(a) shows three typical images of the thin pegs captured below the silo with orifice diameter $D=54$ mm at rotation speeds $\omega$ = 0, 0.48, 0.96 rps, from left to right, respectively. The physical dimensions of these images are  $50 \times 50$ mm$^2$.
The cylindrical wall of the silo is 18 mm from the left edge of the images and the center of the orifice is 27 mm from the right edge. 
The motion of the bottom is in the direction pointing downward in these images.
From these three images, the differences in the alignment of the long axes of the pegs to the motion of the bottom plate at different rotation speeds are not obvious. 
Hence, we developed image analysis codes which take advantage of the grooves along the pegs to measure the orientation angle $\theta$ of a peg relative to the motion of the bottom as shown in the schematic diagram in FIG. \ref{fig:HistThetaAll}(b). 
Then we construct the histograms of $\theta$ using the statistics collected from the image sequences of different rotation speeds.

FIG. \ref{fig:HistThetaAll}(b) displays the histograms $H_\theta$ of a thousand samplings from the image sequences at different rotation speeds for the thin pegs. 
One can see that the orientation angle for zero rotation speed (i.e., $\omega$=0.00 rps) is broadly distributed. 
At finite $\omega$, a peak is observed in the histograms and the width of the distribution represented by the standard deviation $\delta\theta\equiv\sqrt{\langle (\theta-\langle \theta \rangle)^2\rangle}$ is shown in FIG. \ref{fig:MeanTheta}(a).
One can see that $\delta\theta$ reduces from $\approx50^\circ$ at $\omega=0$ rps to $\approx30^\circ$ at  $\omega=1.28$ rps.
Thus, the rotating bottom has an orienting effect for the grains. The average orientation encloses a nonzero angle with the tangential direction of about $\langle\theta\rangle \approx 15^\circ$ for all cases with rotating bottom (see FIG. \ref{fig:MeanTheta}(b)).
We also note that the strength of the orientational ordering does not depend significantly on the actual value of $\omega$.
These observations are in good agreement with earlier experiments on shear flows of similar pegs in a split-bottom Couette cell  \cite{Borzsonyi12,Borzsonyi13}.

\subsection{Effect of orifice position}
\begin{figure}[b]\includegraphics[width=\linewidth]{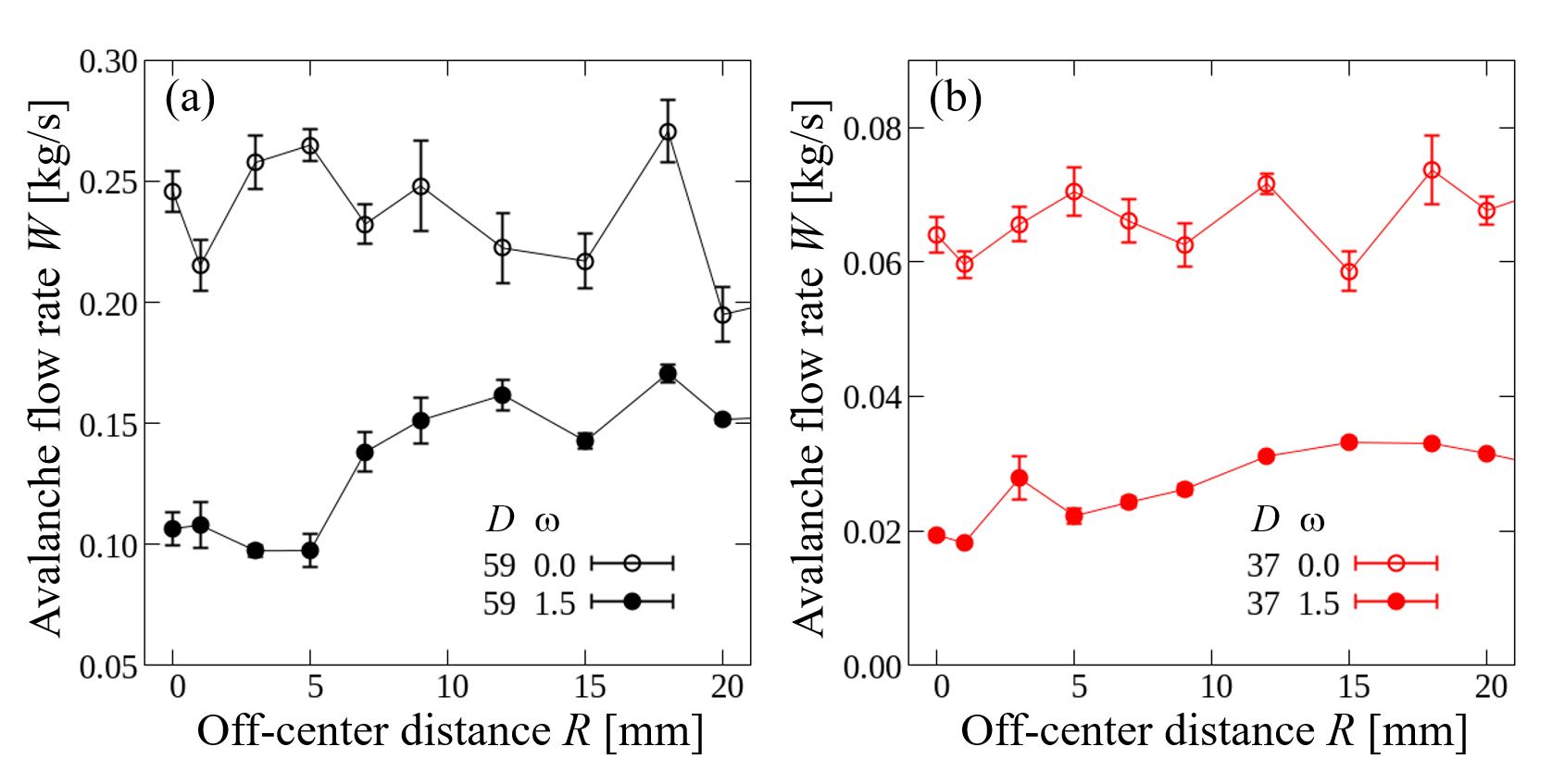}
\caption{Flow rate $W$ vs orifice position $R$ for $D$=59 mm (a) and 37 mm (b) with rotation speed $\omega$ = 0 rps (open circles, stationary) and 1.5 rps (filled circles, fastest).}    \label{fig:R-W}\end{figure}

It is worth mentioning the effects of the distance $R$ of the orifice from the center of the silo.
In principle, if the bottom is not rotating the flow rate should be independent of $R$ as long as the orifice is not too close to the wall of the silo.
FIG. \ref{fig:R-W} shows the avalanche flow rate $W$ plotted against $R$ for the thin pegs obtained for $D$ = 59 mm (a) and 37 mm (b) at $\omega$ = 0.0 rps and 1.5 rps.  
If the bottom is stationary, no significant trend is observed between $W$ and $R$ for both orifice diameters.
When the bottom rotates, for small $R$ the discharge rate $W$ is found to be insensitive to $R$, but increases by about 50\% when $R$ becomes comparable to the particle size. The increase of $W$ with increasing $R$ is coherent with our earlier observations with spherical particles \cite{To19}, but the nature of the curve is slightly different in the two cases. Namely, for spherical particles the flow rate $W$ gradually increased with increasing $R$.
The differences should be related to orientational effects, but further studies are needed to clarify these details.

\section{Summary and conclusion}
To summarize, we report experimental results on the discharge process of elongated grains through a circular orifice in a silo with a rotating bottom.
The conditions in terms of the orifice diameter and the rotation rate for continuous flow, intermittent flow and persistent clog are found. 
If the bottom is stationary, Beverloo law can describe the dependence of avalanche flow rate on orifice diameter all the way from the continuous flow regime to intermittent flow regime down to an orifice diameter of about 3 times the effective grain size of the pegs.
If the bottom rotates with finite speed, the relation between the flow rates and the orifice diameter can be fitted to an exponential function in the intermittent flow regime.
Finite rotation of the bottom plate turns the discharge process to mass flow because pegs at bottom are observed to move to the orifice and contribute to the total discharge current.  
While the current along the bottom slightly enhances the discharge rate in the intermittent flow regime, it reduces the discharge rate in the continuous flow regime.
We also examine the orientation of the pegs at the bottom and we find that the pegs are aligned with a finite angle with the direction of the motion of bottom.
Finally, our preliminary data show that if the bottom rotates, the flow rate can be enhanced if the orifice is placed at a the distance from the center of the bottom. 

To conclude, our experimental results confirm a general feature of silo discharge of spherical or elongated grains---rotating the orifice with respective to the stationary silo wall can switch the dynamics of discharge from funnel flow to mass flow. 
Although the effects of rotation to the flow rates of these two types of grains are similar in general, the reduction of the flow rates in the continuous flow regime for the elongated grains are much larger than that for the spherical grains. This can be explained qualitatively by the alignment of the pegs along the bottom that act like an obstacle to the vertical current through the orifice. A quantitative assessment of the reduction needs further investigation. Also, the observation of the exponential dependence of the flow rates on orifice diameter in the intermittent flow regime is interesting and could be the subject of future investigations. 

\section*{Acknowledgments}
This research is supported by the Ministry of Science and Technology of the Republic of China grants \#: MOST-107-2112-M-001-025, T.B. and T.P. acknowledge support by the EU Horizon 2020 MSCA ITN program CALIPER with Grant No. 812638 and the Hungarian National Research, Development and Innovation Office (NKFIH), under Grant No. OTKA K 116036.

\bibliography{pegs}

\begin{thebibliography}{31}%
\makeatletter
\providecommand \@ifxundefined [1]{%
 \@ifx{#1\undefined}
}%
\providecommand \@ifnum [1]{%
 \ifnum #1\expandafter \@firstoftwo
 \else \expandafter \@secondoftwo
 \fi
}%
\providecommand \@ifx [1]{%
 \ifx #1\expandafter \@firstoftwo
 \else \expandafter \@secondoftwo
 \fi
}%
\providecommand \natexlab [1]{#1}%
\providecommand \enquote  [1]{``#1''}%
\providecommand \bibnamefont  [1]{#1}%
\providecommand \bibfnamefont [1]{#1}%
\providecommand \citenamefont [1]{#1}%
\providecommand \href@noop [0]{\@secondoftwo}%
\providecommand \href [0]{\begingroup \@sanitize@url \@href}%
\providecommand \@href[1]{\@@startlink{#1}\@@href}%
\providecommand \@@href[1]{\endgroup#1\@@endlink}%
\providecommand \@sanitize@url [0]{\catcode `\\12\catcode `\$12\catcode
  `\&12\catcode `\#12\catcode `\^12\catcode `\_12\catcode `\%12\relax}%
\providecommand \@@startlink[1]{}%
\providecommand \@@endlink[0]{}%
\providecommand \url  [0]{\begingroup\@sanitize@url \@url }%
\providecommand \@url [1]{\endgroup\@href {#1}{\urlprefix }}%
\providecommand \urlprefix  [0]{URL }%
\providecommand \Eprint [0]{\href }%
\providecommand \doibase [0]{http://dx.doi.org/}%
\providecommand \selectlanguage [0]{\@gobble}%
\providecommand \bibinfo  [0]{\@secondoftwo}%
\providecommand \bibfield  [0]{\@secondoftwo}%
\providecommand \translation [1]{[#1]}%
\providecommand \BibitemOpen [0]{}%
\providecommand \bibitemStop [0]{}%
\providecommand \bibitemNoStop [0]{.\EOS\space}%
\providecommand \EOS [0]{\spacefactor3000\relax}%
\providecommand \BibitemShut  [1]{\csname bibitem#1\endcsname}%
\let\auto@bib@innerbib\@empty
\bibitem [{\citenamefont {To}\ \emph {et~al.}(2001)\citenamefont {To},
  \citenamefont {Lai},\ and\ \citenamefont {Pak}}]{To01}%
  \BibitemOpen
  \bibfield  {author} {\bibinfo {author} {\bibfnamefont {K.}~\bibnamefont
  {To}}, \bibinfo {author} {\bibfnamefont {P.-Y.}\ \bibnamefont {Lai}}, \ and\
  \bibinfo {author} {\bibfnamefont {H.~K.}\ \bibnamefont {Pak}},\ }\href
  {http://link.aps.org/doi/10.1103/PhysRevLett.86.71} {\bibfield  {journal}
  {\bibinfo  {journal} {Physical Review Letters}\ }\textbf {\bibinfo {volume}
  {86}},\ \bibinfo {pages} {71} (\bibinfo {year} {2001})}\BibitemShut {NoStop}%
\bibitem [{\citenamefont {To}(2005)}]{To05}%
  \BibitemOpen
  \bibfield  {author} {\bibinfo {author} {\bibfnamefont {K.}~\bibnamefont
  {To}},\ }\href {http://link.aps.org/doi/10.1103/PhysRevE.71.060301}
  {\bibfield  {journal} {\bibinfo  {journal} {Physical Review E}\ }\textbf
  {\bibinfo {volume} {71}},\ \bibinfo {pages} {060301(R)} (\bibinfo {year}
  {2005})}\BibitemShut {NoStop}%
\bibitem [{\citenamefont {Zuriguel}\ \emph {et~al.}(2003)\citenamefont
  {Zuriguel}, \citenamefont {Pugnaloni}, \citenamefont {Garcimart\'{i}n},\ and\
  \citenamefont {Maza}}]{Zuriguel03}%
  \BibitemOpen
  \bibfield  {author} {\bibinfo {author} {\bibfnamefont {I.}~\bibnamefont
  {Zuriguel}}, \bibinfo {author} {\bibfnamefont {L.~A.}\ \bibnamefont
  {Pugnaloni}}, \bibinfo {author} {\bibfnamefont {A.}~\bibnamefont
  {Garcimart\'{i}n}}, \ and\ \bibinfo {author} {\bibfnamefont {D.}~\bibnamefont
  {Maza}},\ }\href {http://link.aps.org/doi/10.1103/PhysRevE.68.030301}
  {\bibfield  {journal} {\bibinfo  {journal} {Physical Review E}\ }\textbf
  {\bibinfo {volume} {68}},\ \bibinfo {pages} {030301(R)} (\bibinfo {year}
  {2003})}\BibitemShut {NoStop}%
\bibitem [{\citenamefont {Corwin}(2008)}]{Corwin08}%
  \BibitemOpen
  \bibfield  {author} {\bibinfo {author} {\bibfnamefont {E.~I.}\ \bibnamefont
  {Corwin}},\ }\href {\doibase 10.1103/PhysRevE.77.031308} {\bibfield
  {journal} {\bibinfo  {journal} {Phys. Rev. E}\ }\textbf {\bibinfo {volume}
  {77}},\ \bibinfo {pages} {031308} (\bibinfo {year} {2008})}\BibitemShut
  {NoStop}%
\bibitem [{\citenamefont {Hilton}\ and\ \citenamefont
  {Cleary}(2010)}]{Hilton10}%
  \BibitemOpen
  \bibfield  {author} {\bibinfo {author} {\bibfnamefont {J.}~\bibnamefont
  {Hilton}}\ and\ \bibinfo {author} {\bibfnamefont {P.}~\bibnamefont
  {Cleary}},\ }\href@noop {} {\bibfield  {journal} {\bibinfo  {journal}
  {Physics of Fluids}\ }\textbf {\bibinfo {volume} {22}},\ \bibinfo {pages}
  {071701} (\bibinfo {year} {2010})}\BibitemShut {NoStop}%
\bibitem [{\citenamefont {Zuriguel}\ \emph {et~al.}(2011)\citenamefont
  {Zuriguel}, \citenamefont {Janda}, \citenamefont {Garcimart\'{i}n},
  \citenamefont {Lozano}, \citenamefont {Arevalo},\ and\ \citenamefont
  {Maza}}]{Zuriguel11}%
  \BibitemOpen
  \bibfield  {author} {\bibinfo {author} {\bibfnamefont {I.}~\bibnamefont
  {Zuriguel}}, \bibinfo {author} {\bibfnamefont {A.}~\bibnamefont {Janda}},
  \bibinfo {author} {\bibfnamefont {A.}~\bibnamefont {Garcimart\'{i}n}},
  \bibinfo {author} {\bibfnamefont {C.}~\bibnamefont {Lozano}}, \bibinfo
  {author} {\bibfnamefont {R.}~\bibnamefont {Arevalo}}, \ and\ \bibinfo
  {author} {\bibfnamefont {D.}~\bibnamefont {Maza}},\ }\href
  {http://link.aps.org/doi/10.1103/PhysRevLett.107.278001} {\bibfield
  {journal} {\bibinfo  {journal} {Physical Review Letters}\ }\textbf {\bibinfo
  {volume} {107}},\ \bibinfo {pages} {278001} (\bibinfo {year}
  {2011})}\BibitemShut {NoStop}%
\bibitem [{\citenamefont {Thomas}\ and\ \citenamefont
  {Durian}(2015)}]{Thomas15}%
  \BibitemOpen
  \bibfield  {author} {\bibinfo {author} {\bibfnamefont {C.~C.}\ \bibnamefont
  {Thomas}}\ and\ \bibinfo {author} {\bibfnamefont {D.~J.}\ \bibnamefont
  {Durian}},\ }\href {http://link.aps.org/doi/10.1103/PhysRevLett.114.178001}
  {\bibfield  {journal} {\bibinfo  {journal} {Physical Review Letters}\
  }\textbf {\bibinfo {volume} {114}},\ \bibinfo {pages} {178001} (\bibinfo
  {year} {2015})}\BibitemShut {NoStop}%
\bibitem [{\citenamefont {Nicolas}\ \emph {et~al.}(2018)\citenamefont
  {Nicolas}, \citenamefont {Garcimart\'{i}n},\ and\ \citenamefont
  {Zuriguel}}]{Nicolas18}%
  \BibitemOpen
  \bibfield  {author} {\bibinfo {author} {\bibfnamefont {A.}~\bibnamefont
  {Nicolas}}, \bibinfo {author} {\bibfnamefont {A.}~\bibnamefont
  {Garcimart\'{i}n}}, \ and\ \bibinfo {author} {\bibfnamefont {I.}~\bibnamefont
  {Zuriguel}},\ }\href {\doibase 10.1103/PhysRevLett.120.198002} {\bibfield
  {journal} {\bibinfo  {journal} {Physical Review Letters}\ }\textbf {\bibinfo
  {volume} {120}},\ \bibinfo {pages} {198002} (\bibinfo {year}
  {2018})}\BibitemShut {NoStop}%
\bibitem [{\citenamefont {B\"orzs\"onyi}\ \emph {et~al.}(2012)\citenamefont
  {B\"orzs\"onyi}, \citenamefont {Szab\'o}, \citenamefont {T\"or\"os},
  \citenamefont {Wegner}, \citenamefont {T\"or\"ok}, \citenamefont {Somfai},
  \citenamefont {Bien},\ and\ \citenamefont {Stannarius}}]{Borzsonyi12}%
  \BibitemOpen
  \bibfield  {author} {\bibinfo {author} {\bibfnamefont {T.}~\bibnamefont
  {B\"orzs\"onyi}}, \bibinfo {author} {\bibfnamefont {B.}~\bibnamefont
  {Szab\'o}}, \bibinfo {author} {\bibfnamefont {G.}~\bibnamefont {T\"or\"os}},
  \bibinfo {author} {\bibfnamefont {S.}~\bibnamefont {Wegner}}, \bibinfo
  {author} {\bibfnamefont {J.}~\bibnamefont {T\"or\"ok}}, \bibinfo {author}
  {\bibfnamefont {E.}~\bibnamefont {Somfai}}, \bibinfo {author} {\bibfnamefont
  {T.}~\bibnamefont {Bien}}, \ and\ \bibinfo {author} {\bibfnamefont
  {R.}~\bibnamefont {Stannarius}},\ }\href {\doibase
  10.1103/PhysRevLett.108.228302} {\bibfield  {journal} {\bibinfo  {journal}
  {Physical Review Letters}\ }\textbf {\bibinfo {volume} {108}},\ \bibinfo
  {pages} {228302} (\bibinfo {year} {2012})}\BibitemShut {NoStop}%
\bibitem [{\citenamefont {B\"orzs\"onyi}\ and\ \citenamefont
  {Stannarius}(2013)}]{Borzsonyi13}%
  \BibitemOpen
  \bibfield  {author} {\bibinfo {author} {\bibfnamefont {T.}~\bibnamefont
  {B\"orzs\"onyi}}\ and\ \bibinfo {author} {\bibfnamefont {R.}~\bibnamefont
  {Stannarius}},\ }\href {\doibase 10.1039/C3SM50298H} {\bibfield  {journal}
  {\bibinfo  {journal} {Soft Matter}\ }\textbf {\bibinfo {volume} {9}},\
  \bibinfo {pages} {7401} (\bibinfo {year} {2013})}\BibitemShut {NoStop}%
\bibitem [{\citenamefont {Artoni}\ and\ \citenamefont
  {Richard}(2019)}]{Artoni2019}%
  \BibitemOpen
  \bibfield  {author} {\bibinfo {author} {\bibfnamefont {R.}~\bibnamefont
  {Artoni}}\ and\ \bibinfo {author} {\bibfnamefont {P.}~\bibnamefont
  {Richard}},\ }\href@noop {} {\bibfield  {journal} {\bibinfo  {journal} {Acta
  Mechanica}\ }\textbf {\bibinfo {volume} {230}},\ \bibinfo {pages} {3055}
  (\bibinfo {year} {2019})}\BibitemShut {NoStop}%
\bibitem [{\citenamefont {Campbell}(2011)}]{Campbell2011}%
  \BibitemOpen
  \bibfield  {author} {\bibinfo {author} {\bibfnamefont {C.}~\bibnamefont
  {Campbell}},\ }\href@noop {} {\bibfield  {journal} {\bibinfo  {journal}
  {Physics of Fluids}\ }\textbf {\bibinfo {volume} {23}},\ \bibinfo {pages}
  {013306} (\bibinfo {year} {2011})}\BibitemShut {NoStop}%
\bibitem [{\citenamefont {Mandal}\ and\ \citenamefont
  {Khakhar}(2016)}]{Mandal2016}%
  \BibitemOpen
  \bibfield  {author} {\bibinfo {author} {\bibfnamefont {S.}~\bibnamefont
  {Mandal}}\ and\ \bibinfo {author} {\bibfnamefont {D.~V.}\ \bibnamefont
  {Khakhar}},\ }\href@noop {} {\bibfield  {journal} {\bibinfo  {journal}
  {Physics of Fluids}\ }\textbf {\bibinfo {volume} {28}},\ \bibinfo {pages}
  {103301} (\bibinfo {year} {2016})}\BibitemShut {NoStop}%
\bibitem [{\citenamefont {Nagy}\ \emph {et~al.}(2017)\citenamefont {Nagy},
  \citenamefont {Somfai}, \citenamefont {Börzsönyi},\ and\ \citenamefont
  {Claudin}}]{Nagy2017}%
  \BibitemOpen
  \bibfield  {author} {\bibinfo {author} {\bibfnamefont {D.~B.}\ \bibnamefont
  {Nagy}}, \bibinfo {author} {\bibfnamefont {E.}~\bibnamefont {Somfai}},
  \bibinfo {author} {\bibfnamefont {T.}~\bibnamefont {Börzsönyi}}, \ and\
  \bibinfo {author} {\bibfnamefont {P.}~\bibnamefont {Claudin}},\ }\href@noop
  {} {\bibfield  {journal} {\bibinfo  {journal} {Physical Review E}\ }\textbf
  {\bibinfo {volume} {96}},\ \bibinfo {pages} {062903} (\bibinfo {year}
  {2017})}\BibitemShut {NoStop}%
\bibitem [{\citenamefont {Trulsson}(2018)}]{Trulsson2018}%
  \BibitemOpen
  \bibfield  {author} {\bibinfo {author} {\bibfnamefont {M.}~\bibnamefont
  {Trulsson}},\ }\href@noop {} {\bibfield  {journal} {\bibinfo  {journal}
  {Journal of Fluid Mechanics}\ }\textbf {\bibinfo {volume} {849}},\ \bibinfo
  {pages} {718} (\bibinfo {year} {2018})}\BibitemShut {NoStop}%
\bibitem [{\citenamefont {Reddy}\ \emph {et~al.}(2009)\citenamefont {Reddy},
  \citenamefont {Kumaran},\ and\ \citenamefont {Talbot}}]{Reddy2009}%
  \BibitemOpen
  \bibfield  {author} {\bibinfo {author} {\bibfnamefont {K.~A.}\ \bibnamefont
  {Reddy}}, \bibinfo {author} {\bibfnamefont {V.}~\bibnamefont {Kumaran}}, \
  and\ \bibinfo {author} {\bibfnamefont {J.}~\bibnamefont {Talbot}},\
  }\href@noop {} {\bibfield  {journal} {\bibinfo  {journal} {Physical Review
  E}\ }\textbf {\bibinfo {volume} {80}},\ \bibinfo {pages} {031304} (\bibinfo
  {year} {2009})}\BibitemShut {NoStop}%
\bibitem [{\citenamefont {B\"orzs\"onyi}\ \emph {et~al.}(2016)\citenamefont
  {B\"orzs\"onyi}, \citenamefont {Somfai}, \citenamefont {Szab\'o},
  \citenamefont {Wegner}, \citenamefont {Mier}, \citenamefont {Rose},\ and\
  \citenamefont {Stannarius}}]{Borzsonyi16}%
  \BibitemOpen
  \bibfield  {author} {\bibinfo {author} {\bibfnamefont {T.}~\bibnamefont
  {B\"orzs\"onyi}}, \bibinfo {author} {\bibfnamefont {E.}~\bibnamefont
  {Somfai}}, \bibinfo {author} {\bibfnamefont {B.}~\bibnamefont {Szab\'o}},
  \bibinfo {author} {\bibfnamefont {S.}~\bibnamefont {Wegner}}, \bibinfo
  {author} {\bibfnamefont {P.}~\bibnamefont {Mier}}, \bibinfo {author}
  {\bibfnamefont {G.}~\bibnamefont {Rose}}, \ and\ \bibinfo {author}
  {\bibfnamefont {R.}~\bibnamefont {Stannarius}},\ }\href {\doibase
  10.1088/1367-2630/18/9/093017} {\bibfield  {journal} {\bibinfo  {journal}
  {New Journal of Physics}\ }\textbf {\bibinfo {volume} {18}},\ \bibinfo
  {pages} {093017} (\bibinfo {year} {2016})}\BibitemShut {NoStop}%
\bibitem [{\citenamefont {Tang}\ and\ \citenamefont
  {Behringer}(2016)}]{Tang16}%
  \BibitemOpen
  \bibfield  {author} {\bibinfo {author} {\bibfnamefont {J.}~\bibnamefont
  {Tang}}\ and\ \bibinfo {author} {\bibfnamefont {R.~P.}\ \bibnamefont
  {Behringer}},\ }\href {\doibase 10.1209/0295-5075/114/34002} {\bibfield
  {journal} {\bibinfo  {journal} {EPL (Europhysics Letters)}\ }\textbf
  {\bibinfo {volume} {114}},\ \bibinfo {pages} {34002} (\bibinfo {year}
  {2016})}\BibitemShut {NoStop}%
\bibitem [{\citenamefont {Ashour}\ \emph {et~al.}(2017)\citenamefont {Ashour},
  \citenamefont {Wegner}, \citenamefont {Trittel}, \citenamefont
  {B\"orzs\"onyi},\ and\ \citenamefont {Stannarius}}]{Ashour17}%
  \BibitemOpen
  \bibfield  {author} {\bibinfo {author} {\bibfnamefont {A.}~\bibnamefont
  {Ashour}}, \bibinfo {author} {\bibfnamefont {S.}~\bibnamefont {Wegner}},
  \bibinfo {author} {\bibfnamefont {T.}~\bibnamefont {Trittel}}, \bibinfo
  {author} {\bibfnamefont {T.}~\bibnamefont {B\"orzs\"onyi}}, \ and\ \bibinfo
  {author} {\bibfnamefont {R.}~\bibnamefont {Stannarius}},\ }\href {<Go to
  ISI>://WOS:000395374600009} {\bibfield  {journal} {\bibinfo  {journal} {Soft
  Matter}\ }\textbf {\bibinfo {volume} {13}},\ \bibinfo {pages} {402} (\bibinfo
  {year} {2017})}\BibitemShut {NoStop}%
\bibitem [{\citenamefont {Szab\'o}\ \emph {et~al.}(2018)\citenamefont
  {Szab\'o}, \citenamefont {Kov\'acs}, \citenamefont {Wegner}, \citenamefont
  {Ashour}, \citenamefont {Fischer}, \citenamefont {Stannarius},\ and\
  \citenamefont {B\"orzs\"onyi}}]{Szabo18}%
  \BibitemOpen
  \bibfield  {author} {\bibinfo {author} {\bibfnamefont {B.}~\bibnamefont
  {Szab\'o}}, \bibinfo {author} {\bibfnamefont {Z.}~\bibnamefont {Kov\'acs}},
  \bibinfo {author} {\bibfnamefont {S.}~\bibnamefont {Wegner}}, \bibinfo
  {author} {\bibfnamefont {A.}~\bibnamefont {Ashour}}, \bibinfo {author}
  {\bibfnamefont {D.}~\bibnamefont {Fischer}}, \bibinfo {author} {\bibfnamefont
  {R.}~\bibnamefont {Stannarius}}, \ and\ \bibinfo {author} {\bibfnamefont
  {T.}~\bibnamefont {B\"orzs\"onyi}},\ }\href {\doibase
  10.1103/PhysRevE.97.062904} {\bibfield  {journal} {\bibinfo  {journal}
  {Physical Review E}\ }\textbf {\bibinfo {volume} {97}},\ \bibinfo {pages}
  {062904} (\bibinfo {year} {2018})}\BibitemShut {NoStop}%
\bibitem [{\citenamefont {Vamsi Krishna~Reddy}\ \emph
  {et~al.}(2018)\citenamefont {Vamsi Krishna~Reddy}, \citenamefont {Kumar},
  \citenamefont {Anki~Reddy},\ and\ \citenamefont {Talbot}}]{Vamsi18}%
  \BibitemOpen
  \bibfield  {author} {\bibinfo {author} {\bibfnamefont {A.}~\bibnamefont
  {Vamsi Krishna~Reddy}}, \bibinfo {author} {\bibfnamefont {S.}~\bibnamefont
  {Kumar}}, \bibinfo {author} {\bibfnamefont {K.}~\bibnamefont {Anki~Reddy}}, \
  and\ \bibinfo {author} {\bibfnamefont {J.}~\bibnamefont {Talbot}},\ }\href
  {\doibase 10.1103/PhysRevE.98.022904} {\bibfield  {journal} {\bibinfo
  {journal} {Physical Review E}\ }\textbf {\bibinfo {volume} {98}},\ \bibinfo
  {pages} {022904} (\bibinfo {year} {2018})}\BibitemShut {NoStop}%
\bibitem [{\citenamefont {Hidalgo}\ \emph {et~al.}(2018)\citenamefont
  {Hidalgo}, \citenamefont {Szabó}, \citenamefont {Gillemot}, \citenamefont
  {Börzsönyi},\ and\ \citenamefont {Weinhart}}]{Hidalgo2018}%
  \BibitemOpen
  \bibfield  {author} {\bibinfo {author} {\bibfnamefont {R.~C.}\ \bibnamefont
  {Hidalgo}}, \bibinfo {author} {\bibfnamefont {B.}~\bibnamefont {Szabó}},
  \bibinfo {author} {\bibfnamefont {K.}~\bibnamefont {Gillemot}}, \bibinfo
  {author} {\bibfnamefont {T.}~\bibnamefont {Börzsönyi}}, \ and\ \bibinfo
  {author} {\bibfnamefont {T.}~\bibnamefont {Weinhart}},\ }\href@noop {}
  {\bibfield  {journal} {\bibinfo  {journal} {Physical Review Fluids}\ }\textbf
  {\bibinfo {volume} {3}},\ \bibinfo {pages} {074301} (\bibinfo {year}
  {2018})}\BibitemShut {NoStop}%
\bibitem [{\citenamefont {Azéma}\ \emph {et~al.}(2012)\citenamefont {Azéma},
  \citenamefont {Descantes}, \citenamefont {Roquet}, \citenamefont {Roux},\
  and\ \citenamefont {Chevoir}}]{Azema2012}%
  \BibitemOpen
  \bibfield  {author} {\bibinfo {author} {\bibfnamefont {E.}~\bibnamefont
  {Azéma}}, \bibinfo {author} {\bibfnamefont {Y.}~\bibnamefont {Descantes}},
  \bibinfo {author} {\bibfnamefont {N.}~\bibnamefont {Roquet}}, \bibinfo
  {author} {\bibfnamefont {J.~N.}\ \bibnamefont {Roux}}, \ and\ \bibinfo
  {author} {\bibfnamefont {F.}~\bibnamefont {Chevoir}},\ }\href@noop {}
  {\bibfield  {journal} {\bibinfo  {journal} {Physical Review E}\ }\textbf
  {\bibinfo {volume} {86}},\ \bibinfo {pages} {031303} (\bibinfo {year}
  {2012})}\BibitemShut {NoStop}%
\bibitem [{\citenamefont {Mandal}\ and\ \citenamefont
  {Khakhar}(2017)}]{Mandal2017}%
  \BibitemOpen
  \bibfield  {author} {\bibinfo {author} {\bibfnamefont {S.}~\bibnamefont
  {Mandal}}\ and\ \bibinfo {author} {\bibfnamefont {D.~V.}\ \bibnamefont
  {Khakhar}},\ }\href@noop {} {\bibfield  {journal} {\bibinfo  {journal} {AIChE
  Journal}\ }\textbf {\bibinfo {volume} {63}},\ \bibinfo {pages} {4307}
  (\bibinfo {year} {2017})}\BibitemShut {NoStop}%
\bibitem [{\citenamefont {To}\ \emph {et~al.}(2019)\citenamefont {To},
  \citenamefont {Yen}, \citenamefont {Mo},\ and\ \citenamefont {Huang}}]{To19}%
  \BibitemOpen
  \bibfield  {author} {\bibinfo {author} {\bibfnamefont {K.}~\bibnamefont
  {To}}, \bibinfo {author} {\bibfnamefont {Y.}~\bibnamefont {Yen}}, \bibinfo
  {author} {\bibfnamefont {Y.-K.}\ \bibnamefont {Mo}}, \ and\ \bibinfo {author}
  {\bibfnamefont {J.-R.}\ \bibnamefont {Huang}},\ }\href {\doibase
  10.1103/PhysRevE.100.012906} {\bibfield  {journal} {\bibinfo  {journal}
  {Physical Review E}\ }\textbf {\bibinfo {volume} {100}},\ \bibinfo {pages}
  {012906} (\bibinfo {year} {2019})}\BibitemShut {NoStop}%
\bibitem [{\citenamefont {To}\ and\ \citenamefont {Tai}(2017)}]{To17}%
  \BibitemOpen
  \bibfield  {author} {\bibinfo {author} {\bibfnamefont {K.}~\bibnamefont
  {To}}\ and\ \bibinfo {author} {\bibfnamefont {H.-T.}\ \bibnamefont {Tai}},\
  }\href {https://link.aps.org/doi/10.1103/PhysRevE.96.032906} {\bibfield
  {journal} {\bibinfo  {journal} {Physical Review E}\ }\textbf {\bibinfo
  {volume} {96}},\ \bibinfo {pages} {032906} (\bibinfo {year}
  {2017})}\BibitemShut {NoStop}%
\bibitem [{\citenamefont {Hernández-Delfin}\ \emph {et~al.}(2020)\citenamefont
  {Hernández-Delfin}, \citenamefont {Pongó}, \citenamefont {To},
  \citenamefont {Börzsönyi},\ and\ \citenamefont {Hidalgo}}]{Delfin20}%
  \BibitemOpen
  \bibfield  {author} {\bibinfo {author} {\bibfnamefont {D.}~\bibnamefont
  {Hernández-Delfin}}, \bibinfo {author} {\bibfnamefont {T.}~\bibnamefont
  {Pongó}}, \bibinfo {author} {\bibfnamefont {K.}~\bibnamefont {To}}, \bibinfo
  {author} {\bibfnamefont {T.}~\bibnamefont {Börzsönyi}}, \ and\ \bibinfo
  {author} {\bibfnamefont {R.~C.}\ \bibnamefont {Hidalgo}},\ }\href@noop {}
  {\bibfield  {journal} {\bibinfo  {journal} {Physical Review E}\ }\textbf
  {\bibinfo {volume} {102}},\ \bibinfo {pages} {042902} (\bibinfo {year}
  {2020})}\BibitemShut {NoStop}%
\bibitem [{\citenamefont {Thomas}\ and\ \citenamefont
  {Durian}(2013)}]{Thomas13}%
  \BibitemOpen
  \bibfield  {author} {\bibinfo {author} {\bibfnamefont {C.~C.}\ \bibnamefont
  {Thomas}}\ and\ \bibinfo {author} {\bibfnamefont {D.~J.}\ \bibnamefont
  {Durian}},\ }\href {http://link.aps.org/doi/10.1103/PhysRevE.87.052201}
  {\bibfield  {journal} {\bibinfo  {journal} {Physical Review E}\ }\textbf
  {\bibinfo {volume} {87}},\ \bibinfo {pages} {052201} (\bibinfo {year}
  {2013})}\BibitemShut {NoStop}%
\bibitem [{\citenamefont {Janda}\ \emph {et~al.}(2008)\citenamefont {Janda},
  \citenamefont {Zuriguel}, \citenamefont {Garcimart\'{i}n}, \citenamefont
  {Pugnaloni},\ and\ \citenamefont {Maza}}]{Janda08}%
  \BibitemOpen
  \bibfield  {author} {\bibinfo {author} {\bibfnamefont {A.}~\bibnamefont
  {Janda}}, \bibinfo {author} {\bibfnamefont {I.}~\bibnamefont {Zuriguel}},
  \bibinfo {author} {\bibfnamefont {A.}~\bibnamefont {Garcimart\'{i}n}},
  \bibinfo {author} {\bibfnamefont {L.~A.}\ \bibnamefont {Pugnaloni}}, \ and\
  \bibinfo {author} {\bibfnamefont {D.}~\bibnamefont {Maza}},\ }\href
  {http://stacks.iop.org/0295-5075/84/i=4/a=44002} {\bibfield  {journal}
  {\bibinfo  {journal} {EPL (Europhysics Letters)}\ }\textbf {\bibinfo {volume}
  {84}},\ \bibinfo {pages} {44002} (\bibinfo {year} {2008})}\BibitemShut
  {NoStop}%
\bibitem [{\citenamefont {Mankoc}\ \emph {et~al.}(2007)\citenamefont {Mankoc},
  \citenamefont {Janda}, \citenamefont {Arévalo}, \citenamefont {Pastor},
  \citenamefont {Zuriguel}, \citenamefont {Garcimart\'{i}n},\ and\
  \citenamefont {Maza}}]{Mankoc07}%
  \BibitemOpen
  \bibfield  {author} {\bibinfo {author} {\bibfnamefont {C.}~\bibnamefont
  {Mankoc}}, \bibinfo {author} {\bibfnamefont {A.}~\bibnamefont {Janda}},
  \bibinfo {author} {\bibfnamefont {R.}~\bibnamefont {Arévalo}}, \bibinfo
  {author} {\bibfnamefont {J.~M.}\ \bibnamefont {Pastor}}, \bibinfo {author}
  {\bibfnamefont {I.}~\bibnamefont {Zuriguel}}, \bibinfo {author}
  {\bibfnamefont {A.}~\bibnamefont {Garcimart\'{i}n}}, \ and\ \bibinfo {author}
  {\bibfnamefont {D.}~\bibnamefont {Maza}},\ }\href {\doibase
  10.1007/s10035-007-0062-2} {\bibfield  {journal} {\bibinfo  {journal}
  {Granular Matter}\ }\textbf {\bibinfo {volume} {9}},\ \bibinfo {pages} {407}
  (\bibinfo {year} {2007})}\BibitemShut {NoStop}%
\bibitem [{\citenamefont {Janda}\ \emph {et~al.}(2012)\citenamefont {Janda},
  \citenamefont {Zuriguel},\ and\ \citenamefont {Maza}}]{Janda12}%
  \BibitemOpen
  \bibfield  {author} {\bibinfo {author} {\bibfnamefont {A.}~\bibnamefont
  {Janda}}, \bibinfo {author} {\bibfnamefont {I.}~\bibnamefont {Zuriguel}}, \
  and\ \bibinfo {author} {\bibfnamefont {D.}~\bibnamefont {Maza}},\ }\href
  {http://link.aps.org/doi/10.1103/PhysRevLett.108.248001} {\bibfield
  {journal} {\bibinfo  {journal} {Physical Review Letters}\ }\textbf {\bibinfo
  {volume} {108}},\ \bibinfo {pages} {248001} (\bibinfo {year}
  {2012})}\BibitemShut {NoStop}%
\end{thebibliography}%

\end{document}